\documentclass[aps,prl,reprint,groupedaddress]{revtex4-1}
\usepackage{graphicx}
\usepackage{siunitx}
\usepackage{afterpage}
\usepackage{subcaption} 

\newdimen\figrasterwd
\figrasterwd\textwidth

\let\oldcite\cite
\renewcommand{\cite}[1]{\mbox{\oldcite{#1}}}

\begin{document}

\title{Fluorescent thermal imaging of a non-insulated pancake coil wound from high temperature superconductor tape}

\author{R. Gyur\'{a}ki}

\affiliation{The authors are with the Karlsruhe Institute of Technology, Eggenstein-Leopoldshafen, Germany}

\author{T. Benkel}
\affiliation{The authors are with the Karlsruhe Institute of Technology, Eggenstein-Leopoldshafen, Germany}

\author{F. Schreiner}
\affiliation{The authors are with the Karlsruhe Institute of Technology, Eggenstein-Leopoldshafen, Germany}

\author{F. Sirois}
\affiliation{Author is with Polytechnique Montr\'{e}al, Montr\'{e}al (QC), Canada}

\author{F. Grilli}
\affiliation{The authors are with the Karlsruhe Institute of Technology, Eggenstein-Leopoldshafen, Germany}

\date{\today}

\begin{abstract}
We have wound a 157-turn, non-insulated pancake coil with an outer diameter of \SI{85}{\milli\meter} and we cooled it down to \SI{77}{\kelvin} with a combination of conduction and gas cooling. Using high-speed fluorescent thermal imaging in combination with electrical measurements we have investigated the coil under load, including various ramping tests and over-current experiments. We have found found that the coil does not heat up measurably when being ramped to below its critical current. Two over-current experiments are presented, where in one case the coil recovered by itself and in another case a thermal runaway occurred. We have recorded heating in the bulk of the windings due to local defects, however the coil remained cryostable even during some over-critical conditions and heated only to about \SI{82}{}--\SI{85}{\kelvin} at certain positions. A thermal runaway was observed at the center, where the highest magnetic field and a resistive joint create a natural defect. The maximum temperature, $\sim$\SI{100}{\kelvin}, was reached only in the few innermost windings around the coil former.
\end{abstract}

\maketitle

Coils wound from second generation high temperature superconductors (HTS) are being researched for use in high field magnets \citep{Weijers2010, Jang2017}, as well as in electrical motors and generators \citep{Nick2012}. Insulated coils -- in which the coil is either co-wound with an electrical insulator or the conductor itself has an insulator coating -- suffer from stability issues and are prone to damage due to local defects. Non-insulated (NI) coils \citep{Hahn2011}, where the HTS tapes are wound without turn-to-turn insulation, are also actively researched \citep{Zhang2017a,Wang2017, Suetomi2016, Wang2015, Wang2015a, Kim2015, Choi2012a} as an alternative. One of the main benefits of such coils is their proven self-protecting behavior \citep{Hahn2011, XWang2013, Wang2016} due alternative current paths via the turn-to-turn resistances.
Our group has previously reported on using a high-speed fluorescent thermal imaging method, where with the help of a fluorescent, temperature-sensitive coating and a high-speed camera, the surface temperature of HTS tapes can be mapped during a quench \citep{Gyuraki2018}. By recording the light intensity changes emitted by the fluorescent coating at high speeds, the development of a hot spot can be closely observed at sub-millisecond temporal resolution while operating at cryogenic temperatures. 
In order to show the possible applications of the method and also to gain more insight about the behavior and thermal properties of NI coils, a small pancake coil was dry-wound without turn-to-run insulation. The aim of the research was to observe where heating occurs in a coil during normal and over-current conditions, without introducing any artificial defect. Two experiments are shown in this publication, in which the coil was driven in over-current conditions and the thermal and electrical behavior was analyzed.

The fluorescent thermal imaging's obvious requirement is that a surface has to be optically visible. This puts significant constraints on both the coil's and the cryostat's design. First, the boiling of a liquid nitrogen bath must be avoided and the coil surface cannot be obstructed by any kind of resin impregnation, current lead or other structural part. In addition, ideally the pancake coil should have a large inner-to-outer diameter distance (winding thickness), in order to allow the observation of any meaningful thermal effect. To solve these problems, a simple conduction plus gas cooled experimental setup was constructed.

A coil was wound from 4-mm wide SuperPower SCS4050 AP HTS coated conductor using a tension of $\sim$\SI{2}{\kilo\gram}, starting with a soldered In-Ag joint on a 30-mm diameter copper former. A separate, 12-mm wide HTS tape was also soldered to the bottom of this copper cylinder. This tape runs at the bottom of the coil and is used as current lead. A total length of approximately \SI{30}{\meter} of HTS tape was used for 157 turns, the outermost two of which were soldered together for providing mechanical stability to the coil. After the soldered outer turns, an extra length of tape of \SI{20}{\centi\meter} was kept for being used as the second current lead.

\begin{figure}
\includegraphics[width=6.5cm]{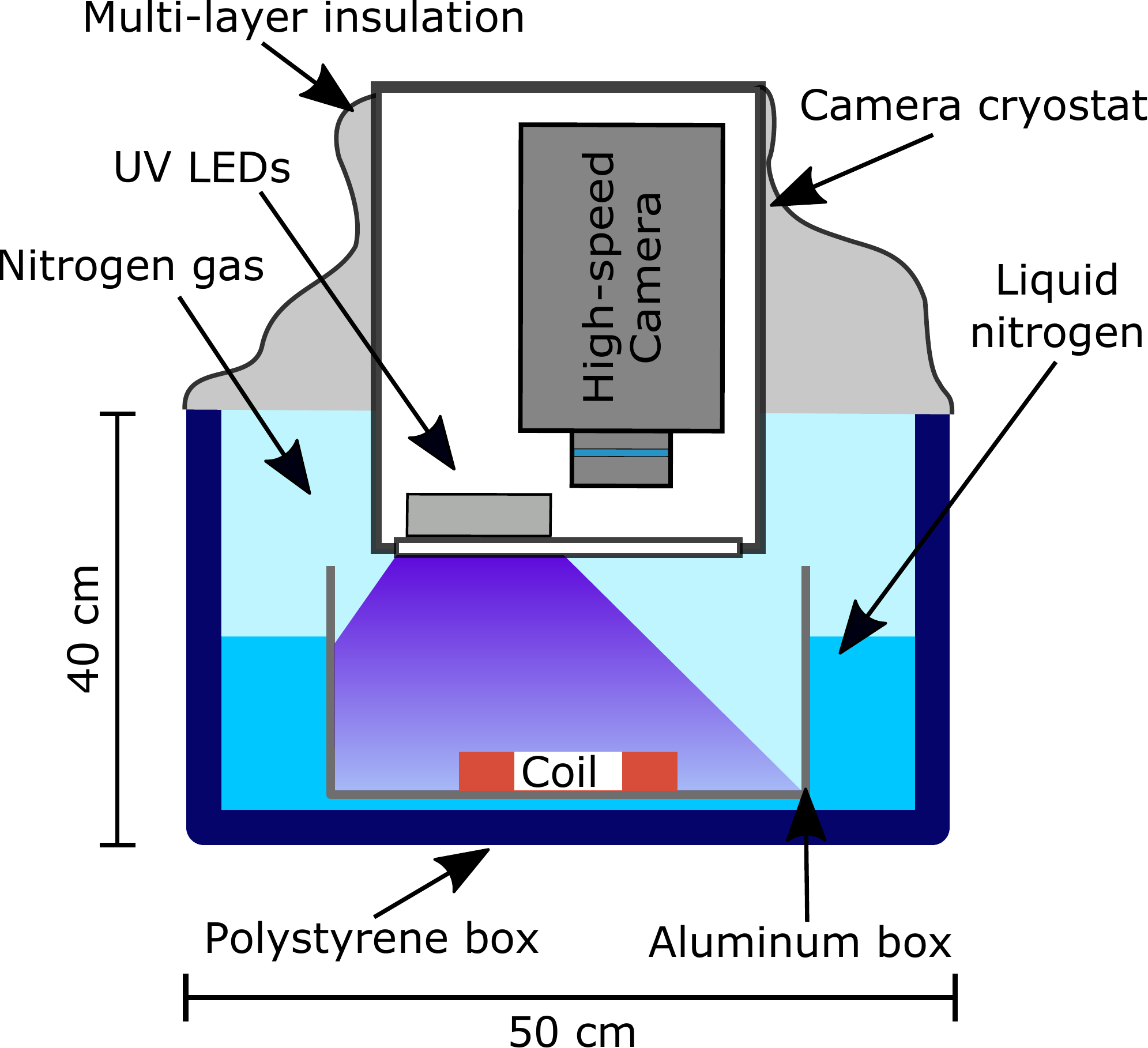}
\caption{\label{exparimental_setup}Cross-section view of the experimental setup.}
\end{figure}

To avoid complex and expensive vacuum and cryocooler setups with optical windows -- and since the aim was to test the coil only at \SI{77}{\kelvin} -- a simple, 2-component cryostat was constructed, as shown in Figure \ref{exparimental_setup}. An aluminum container was placed inside a larger polystyrene container, and the space between the two was filled up with liquid nitrogen. The coil was then placed into the aluminum box and a good thermal contact was made between the two with the help of a thin layer of thermal paste. This allowed the coil to be conduction-cooled to \SI{77}{\kelvin} via the aluminum container's walls. The remaining space in the containers was filled with evaporated, cold nitrogen gas, providing a cold and clear atmosphere, ideal for optical measurements. The high-speed camera was then positioned above the setup and lowered close to the coil inside a separate optical cryostat that kept the camera at room temperature. The UV excitation light was also placed in the camera's cryostat. The setup was finally surrounded by multi-layer insulation from the top to shield the sample from radiation heat sources, and to prevent air movement and mixing with the cold nitrogen gas, and also to block out ambient light. The result was a conduction-cooled NI-coil in a pure, cold, nitrogen atmosphere.
The high-speed camera, despite having been used in the past at several thousand frames per second (fps) to measure quenches in single HTS tapes \citep{Gyuraki2018}, was used here at the slowest speed setting of 50 fps, which was found sufficient for measuring the heating in the NI coil. This allowed a maximum recording time of about 10-12 minutes per experiment.

To observe local heating over the coil's surface, the same droplet coating method was used as in experiments with HTS tapes described in \citep{Gyuraki2018}.

\begin{figure}[b]
\includegraphics[width=80mm]{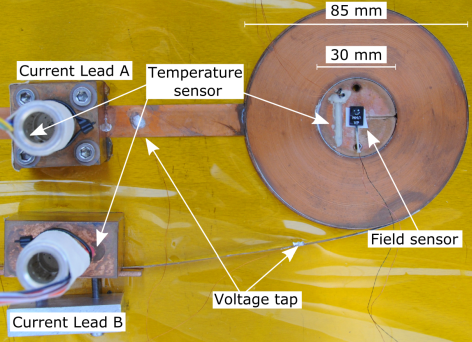}
\caption{\label{coil_picture} HTS non-insulated coil placed at the bottom of an aluminium box. A Kapton sheet served as electrical insulation in between.}
\end{figure}

A picture of the coil and the electrical instrumentation is shown in Figure \ref{coil_picture}. The magnetic field was measured with a Hall sensor and the temperature with a Pt100 thermometer, both located at the center of the coil. The voltage was measured across two superconducting current leads. However, this path includes the resistive soldered joint between the 12-mm HTS current lead and the coil former, as well as between the coil former and the first turn. Since the current leads were also conduction-cooled, their temperatures were also recorded with a Pt100 sensor each. The current was measured with a shunt resistor inserted in the circuit. This resulted in seven data acquisition channels, which were recorded using an NI-USB 6289 DAQ, at a rate of 1000 samples per second, and several Dataforth and Dewetron pre-amplifiers for acquiring low voltages. In parallel, the coil voltage was also measured with a nanovoltmeter at a slower rate of 1 sample per second for higher accuracy.

\setcounter{figure}{3}
\begin{figure*}[tb]
\begin{center}
\vspace*{\fill}
  \parbox{\figrasterwd}{
    \parbox{0.282\textwidth}{%
      \subcaptionbox{t=297.8 s\label{thermal_72A_OC_a}}{\includegraphics[width=54mm]{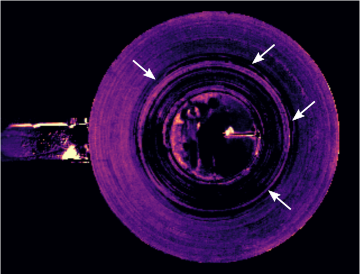}}
    }
    \hskip1em
    \parbox{0.282\textwidth}{%
      \subcaptionbox{t=298.8 s\label{thermal_72A_OC_b}}{\includegraphics[width=54mm]{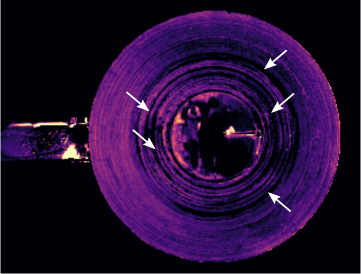}}
    }
    \hskip1em
    \parbox{0.388\textwidth}{%
      \subcaptionbox{t=300 s \label{thermal_72A_OC_c}}{\includegraphics[width=72mm]{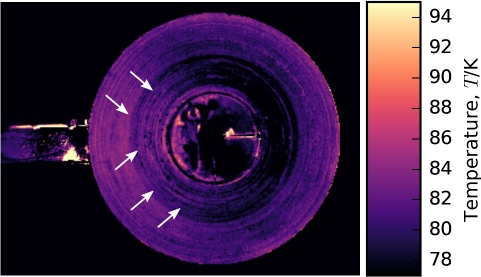}
      }
    }
  }
\caption{\label{thermal_72A_OC} Thermal images of the coil during the \SI{72}{\ampere} over-current test showing different time points just before and right after the ``self-recovery''. The white arrows indicate features, as described in the text.}
\end{center}
\vfill
\end{figure*}

\setcounter{figure}{2}
\begin{figure}[b]
\includegraphics[width=88mm]{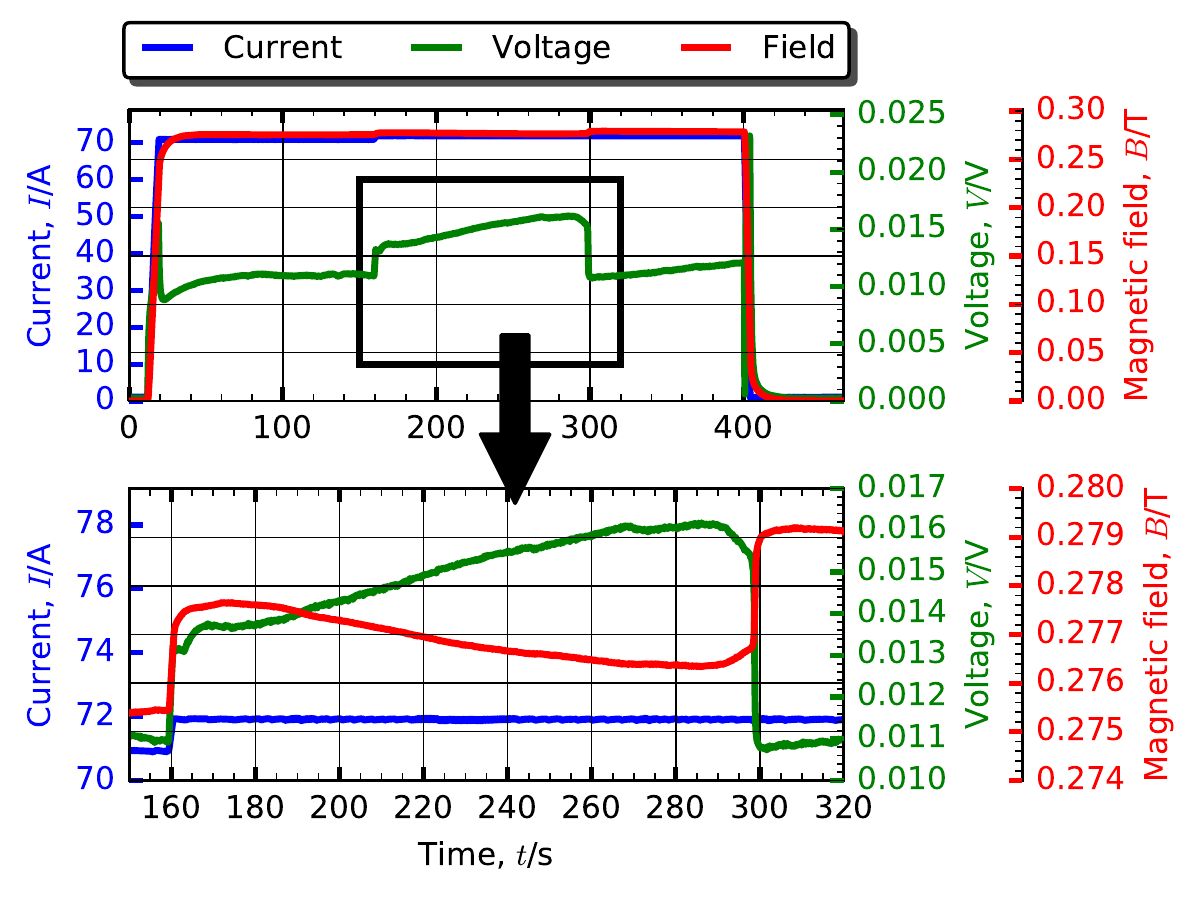}
\caption{\label{OC_72A}Electrical measurements of the coil during the \SI{72}{\ampere} over-current test. The bottom plot shows a zoomed area marked by the black rectangle. Note that the axes ranges are changed on the zoomed plot.}
\end{figure}

\setcounter{figure}{5}
\begin{figure*}[tb]
\begin{center}
\vspace*{\fill}
  \parbox{\figrasterwd}{
    \parbox{0.272\textwidth}{%
      \subcaptionbox{t=320.0 s\label{thermal_74A_OC_a}}{\includegraphics[width=52.6mm]{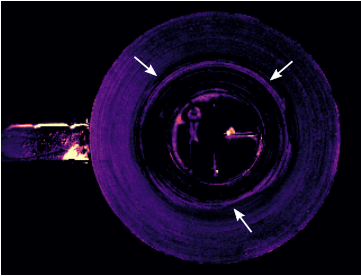}}
    }
    \hskip1em
    \parbox{0.272\textwidth}{%
      \subcaptionbox{t=340.0 s\label{thermal_74A_OC_b}}{\includegraphics[width=52.6mm]{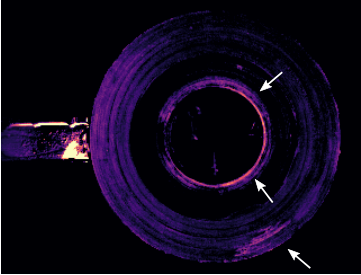}}
    }
    \hskip1em
    \parbox{0.4\textwidth}{%
      \subcaptionbox{t=345.0s\label{thermal_74A_OC_c}}{\includegraphics[width=75mm]{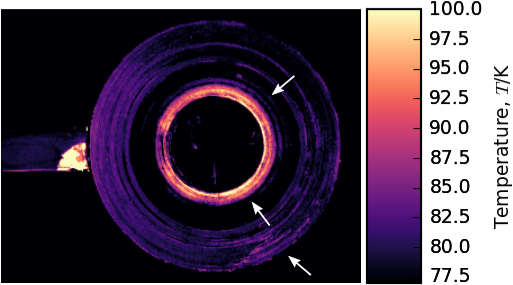}}
\vspace{2px}
    }
  }
\caption{\label{thermal_74A_OC} Thermal images of the coil during the \SI{74}{\ampere} over-current test showing different time points just before and at the moment of thermal runaway and ramping down. The white arrows indicate features, as described in the text.}
\vfill
\end{center}
\end{figure*}

Before testing the coil in the conduction-cooled setup, it was first measured in a liquid nitrogen bath to obtain its critical current using the \SI{0.1}{\micro\volt/\centi\meter} and \SI{1}{\micro\volt/\centi\meter} criteria, which worked out to be \SI{57}{\ampere} and \SI{67}{\ampere}, respectively. 
In the conduction-cooled setup, the coil was ramped using several ramping rates, while being recorded at 50 fps to observe whether the ramping rate caused any heating during ramp-up. The tested ramping rates ranged from \SI{0.5}{\ampere/\second} to \SI{1000}{\ampere/\second}. It was found that virtually no heating took place in the coil during ramping, with the exception of the 12-mm wide HTS current lead. The latter does not sit completely flat on the conduction-cooled surface and hence it does heat up initially. However, being 3 times wider than the tape in the coil, it could easily withstand currents up to \SI{74}{\ampere}.
The absence of any remarkable heating in the test coil even during fast ramps is a consequence of its low critical current, leading to small Joule losses when the current flows across the neighboring coil turns.

In our over-current measurements, the coil was first ramped up to previously validated safe current levels and from there, slowly to higher currents. At specific current levels the ramping was stopped and the coil was held steady to observe if its voltage was still stable. This stepwise ramping was then continued until the coil voltage could no longer stabilize and a ``natural'' runaway appeared. Special care was taken to manually ramp down the coil in case a high voltage arose, in order to permit future experimentation and avoid damages.
In different measurements, the transition point was observed at slightly different current levels, which suggests that slight disturbances in cooling, ramping and current levels can influence the thermal stability to some degree.

The coil current was first ramped up to \SI{71}{\ampere} at a rate of \SI{10}{\ampere/\second} and the voltage was monitored, as shown in Figure \ref{OC_72A}. It eventually stabilized and so the current was increased to \SI{72}{\ampere}, this time at a slower \SI{1}{\ampere/\second} rate, as shown at time 160 s. At this current level, a slow yet steady voltage shift was observed, and the coil was held at this level. 

Since the magnetic field also reduced with the increasing voltage, the current must have been changing from the spiral to the radial path. This is also supported by the quasi-linear voltage rise observed, as expected with increasingly more current flowing through a resistor. In Figure~\ref{thermal_72A_OC_a} one can see that the outer turns of the coil are initially at a higher temperature, presumably due to slightly worse cooling than in other parts of the coil. At the same time, distinct heating is visible in the form of circles with varying width, as can be seen in Figure~\ref{thermal_72A_OC_a} and Figure~\ref{thermal_72A_OC_b} highlighted by the white arrows. This indicates that at particular locations in the coil, the current flows through the turn-to-turn contact resistance and generate losses. At about 298 seconds into the measurement, the voltage dropped and the magnetic field increased sharply, indicating that the current reverted back into the spiral path quite abruptly. Simultaneously, the center windings of the coil have heated up, as indicated by the white arrows in Figure \ref{thermal_72A_OC_c}, where the ring-like hot areas have been replaced by an uniformly warmer zone, mostly on the left hand side of the coil. Hence the current flowing though the turn-to-turn contact resistances of the coil can be seen on the thermal image as heating over possibly several turns, which is most likely a function of the cooling performance and the distribution of local fluctuations of $I_{\rm{c}}$ along the length of the wound conductor.

\setcounter{figure}{4}
\begin{figure}[b]
\includegraphics[width=88mm]{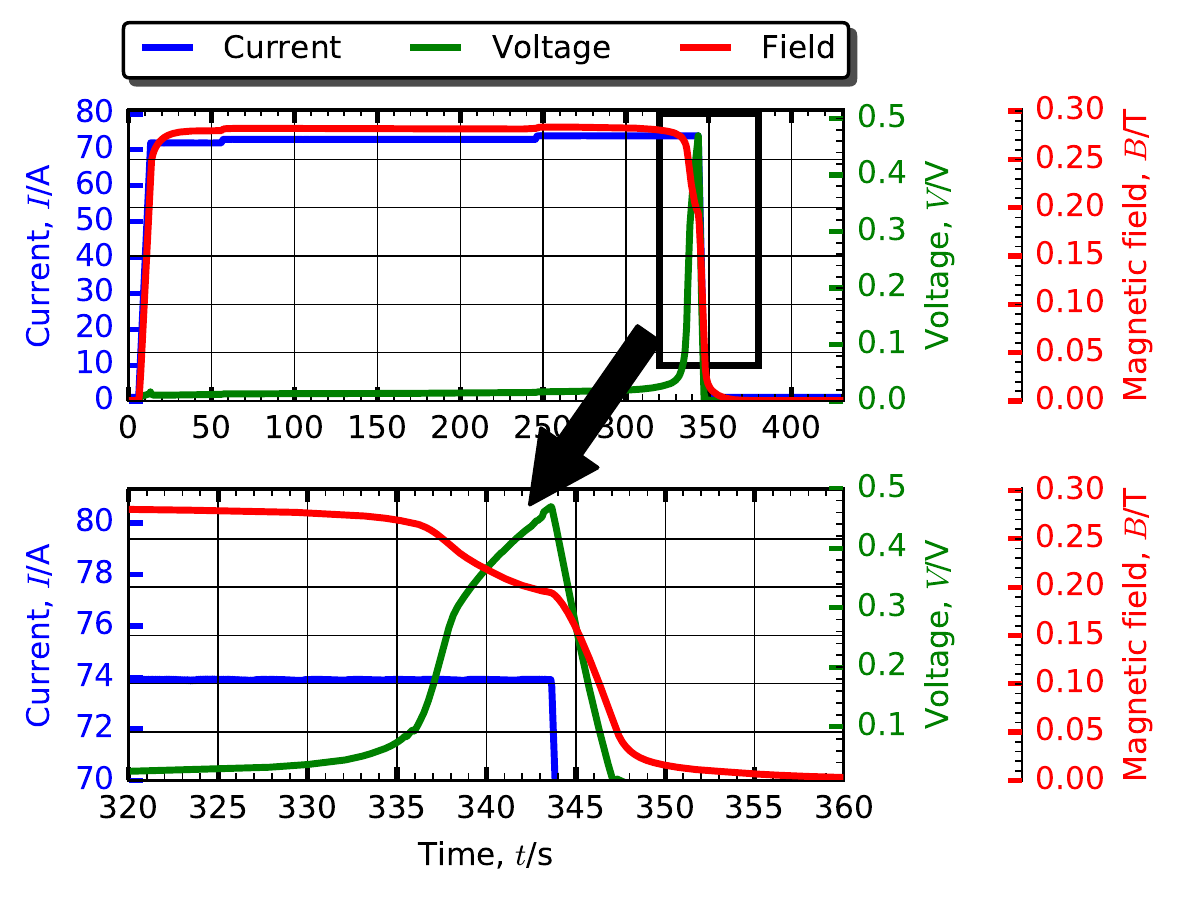}
\caption{\label{OC_74A}Electrical measurements of the coil during the \SI{74}{\ampere} over-current test. The bottom plot shows a zoomed area marked by the black rectangle.}
\end{figure}

To reproduce previous results and observe the same behavior again, the coil current was ramped up in a subsequent measurement directly to \SI{72}{\ampere} at a rate of \SI{10}{\ampere/\second}, as shown in Figure \ref{OC_74A}. However, in this case, no voltage drift was observed. The current was therefore ramped up to \SI{73}{\ampere}, and finally \SI{74}{\ampere}, at a slower \SI{1}{\ampere/\second} rate, where the coil voltage became unstable and started increasing once again. 
Over the course of two minutes at a steady current of \SI{74}{\ampere}, the coil voltage rose up continuously until its final value of \SI{0.47}{\volt}, at which the coil current was ramped down quickly. The field started decaying with the developing voltage, indicating a slow current transfer into a radial path. Then, after \SI{336}{\second}, a much sharper voltage rise and accompanying field drop became visible, likely due to (some of) the HTS tapes becoming resistive.
Figure \ref{thermal_74A_OC} shows excerpts of the thermal imaging recording at three different time instants. In Figure \ref{thermal_74A_OC_a}, one can see that that the outside of the coil is again at a higher temperature, however still cryostable. Warmer zones are still visible around the center and middle layer of the coil (shown by the white arrows), indicating heating due to currents flowing through the turn-to-turn contact resistances. Figure \ref{thermal_74A_OC_b} shows that the middle layers seem to have mostly cooled down, and at the same time -- due to the voltage rise of the coil -- the innermost windings started heating up. At the bottom of the image, where the outer current lead arrives at the coil, a hot spot is also visible. Then in Figure \ref{thermal_74A_OC_c}, just before the current was ramped down to protect the coil, the hottest temperature of around \SI{100}{\kelvin} was reached at the few innermost windings of the coil. When looking at the dynamic development (see attached video \footnote{The attached video shows the thermal imaging as well as the electrical signals. This latter includes the current, magnetic field, voltage across the coil and the temperature measured by the Pt100 sensor as shown in Figure~\ref{coil_picture}}) one can identify a stage between Figure \ref{thermal_74A_OC_a} and Figure \ref{thermal_74A_OC_b}. Between these times the middle turns of the coil actually cooled down, and when the voltage continued to rise, the central windings started heating up instead. A possible explanation is that when the coil voltage is low, only the ``major'' defects in the coil can be bypassed. However, at higher voltages, more current can be forced through the turn-to-turn contact resistances, thereby reducing the current in the superconductor as well as the local heating. Due to the fact that the energy is dissipated over a larger volume, this could then manifest in a form of a local cooling effect. Then in Figure \ref{thermal_74A_OC_c}, the center windings see the combined effect of superconducting to resistive transition of the HTS due to the reduced temperature margin and high magnetic field, combined with the current bypass and joule heating due to the resistive joints.
To check for damages, after each measurement, the coil resistance was measured and compared with its previous values. At the end, a second $I_{\rm{c}}$ measurement was also done and no degradation was observed.

To conclude, the applicability of fluorescent thermal imaging in applied superconductivity was demonstrated on the example of an HTS-wound NI pancake coil. In two over-current measurements, different thermal and electrical behaviors were observed. In the first experiment, the coil current was ramped up until \SI{72}{\ampere} where a slow, resistive voltage shift was observed, together with a linear drop in the magnetic field. However, the magnetic field recovered and the voltage dropped abruptly, which was accompanied by a sudden change in heating distribution over the coil. In a successive measurement, the coil current was ramped up to \SI{74}{\ampere}, where a characteristic superconducting runaway was observed. Here, during the rapid voltage rise, the coil's innermost windings heated up, whereas the remainder of the coil stayed cold. This confirmed that the initial point of the thermal runaway was located at a resistive joint, e.g. the central copper former, coinciding with the highest magnetic field. Another observation was the change in voltage rise during the thermal runaway. Initially, the typical sharp transition signal was observed on the voltage measurement, but soon after, the rise became more linear. The thermal imaging has also shown that current redistribution (e.g. due to local $I_{\rm{c}}$ variations) does not happen at a single point around a local defect, but rather over larger length scales. This is visible in the thermal images as warmer rings spanning over a few turns.

\clearpage
\bibliography{references}

\end{document}